**Parametric amplification of electromagnetic waves produced by a**

**flux-flow-oscillator made of YBa$_2$Cu$_3$O$_{7-\delta}$ Josephson junction arrays**

Boris Chesca[a1], Daniel John[a], Marat Gaifullin[a,b]

[a]Department of Physics, Loughborough University

Loughborough, LE11 3TU, United Kingdom

[b]SuperOx Japan LLC, Sagamihara, 252-0243, Japan

**Abstract**

We observe parametric amplification of electromagnetic (EM) waves produced by a flux-flow oscillator made of YBa$_2$Cu$_3$O$_{7-\delta}$ Josephson junctions arrays coupled to the resonant modes of a millimeter wave Fabry-Perot resonator at a pump frequency $f_P$=45 GHz. For temperatures in the range (30-45) K the frequency $f_S$ of the EM signal to be amplified could be tuned continuously in the range (1-25) GHz by an applied $B$-field induced flux $\Phi$ with a one-flux-quantum $\Phi_0$ periodicity. Consequently, we measured a significant parametric gain that is almost frequency independent, with a maximum of (8-10.4) dB reached at 40K. For temperatures in the range (14-30) K the magnetic field tunability of $f_S$ is gradually suppressed to a minimum of (1-5) GHz range where a parametric gain between (5-6) dB was measured. With an appropriate adjustment of design/fabrication parameters our results suggests that the development of tunable *MW* generators/detectors, as well as parametric amplifiers made of high transition temperature superconductors and operating in a wide range of temperatures (10 mK-77K) is a reasonable and appealing possibility.

Flux-flow-oscillators (FFO) based on unidirectional flow of magnetic vortices in a single-long Josephson junction (JJ) [1-5] are natural, tunable, emitters/receivers of electromagnetic (*EM*) radiation. FFO's operating at 4.2 K have been implemented to develop ultra-sensitive superconducting sub-terahertz integrated-receivers routinely used in radio-astronomical research [6] or atmospheric science projects [7-8]. Developing narrow-band FFO for sub-terahertz integrated-receivers and microwave (*MW*) generators operating at more practical temperatures than 4.2 K, say in the range (30, 77) K, would be an important achievement. Here, we present a significant progress in this direction by reporting on the experimental observation of parametrically-amplified *EM* produced by a FFO made of YBa$_2$Cu$_3$O$_{7-\delta}$ (YBCO) Josephson junctions (*JJs*) array coupled to the resonant modes of a

---

millimeter wave Fabry-Perot resonator. Our results are also particular relevant considering the increased attention to Josephson parametric amplifiers (*JPAs*) firstly introduced decades ago [9-13] due to their applications in quantum technology, in particular, as amplifiers/readouts of superconducting quantum bits [14-22]. Currently, two approaches are used for amplification in superconducting qubits read-out schemes. The first one is based on an all-semiconductor-based technology with a high electron mobility transistor (HEMT) amplifier operated at (3-4) K followed by a room temperature amplifier. The second approach is a hybrid one with a superconducting parametric amplifier made of Nb-Al trilayer technology operated at (10, 20) mK used in the very first stage of amplification followed by a HEMT and a room temperature amplifiers. From this perspective, our report suggests a potentially interesting alternative of using exclusively superconducting technologies for all stages of amplification from mK range (Nb-Al technology in the first stage) to 77 K (YBCO technology in the following stages: *JPAs* and series SQUID arrays [23-24]). Parametric amplifiers operate based on one fundamental principle: the incoming *EM* signal of frequency $f_S = \omega_S/2\pi$ is mixed with an applied *EM* pump tone at $f_P = \omega_P/2\pi$ via an intrinsic nonlinearity, by which energy from the pump is converted into *EM* signal and thereby providing gain. In our case, we designed, fabricated and tested *JPAs* based on *FFOs* made of series array of 100 *JJs* separated by superconducting rectangular loops and parametrically pumped its loop inductances via a Fabry-Perot resonator at a frequency $f_P$ = 45 GHz by modulating the flux threading the array's loops. By comparing the height of the *EM*-induced resonant flux-flow resonances on the array's *dc* current-voltage characteristics (*IVC's*) for a wide range of values in the (*V*, *B*) plane with the microwave (*MW*) pump on and off we calculated a parametric gain in the range (6-10) dB for operation temperatures between 30 K and 45 K and an *EM* signal of frequency $f_S$ which could be tuned by an applied magnetic field in the frequency range (1-25) GHz. For temperatures in the range (14-30) K the magnetic field tunability of $f_S$ is strongly suppressed to the (1-5) GHz range where a parametric gain between (5-6) dB was measured.

It is useful to briefly clarify the physical origin of *EM*-radiation emitted by a *FFO* based on *JJ*-arrays [23-29]. When a *B* field is applied perpendicular to a planar one-dimensional *JJ*-array, magnetic vortices will enter the array in a form of Josephson vortices. The bias current, *I*, flowing across the array produces a Lorentz force which drives the Josephson vortices unidirectionally, forming a lattice of vortices moving with a certain speed. This is accompanied by an emission of EM-waves that propagate along the array. When the vortex spacing is commensurate with the wavelength of emitted EM-waves, resonant modes occur. This can be viewed as the phase-locking condition of the vortex velocity and the phase velocity of one of the self-induced EM-modes. The experimental signature of such phase locking between a train of propagating vortices and their induced *EM*-radiation in a *JJ*-array is a series of flux-flow resonances m=1-4 in the dc current-voltage characteristics (*IVC's*). On a resonant current step, moving vortices couple to their induced linear waves. The emission of the EM-radiation is maximum at the resonance and is typically in the region of hundreds of nW.

The *JJ*-arrays were fabricated by depositing high quality epitaxial, 100 nm thick *c*-axis oriented YBCO films on 10x10 mm$^2$, 24° symmetric [001] tilt SrTiO$_3$ bicrystals by pulsed laser deposition. Medium-underdoped YBCO films with a critical temperature $T_c$=49 K were subsequently patterned by optical lithography and etched by an Ar ion beam to form 100 JJ-arrays (see Fig. 1a). Within the 100 JJ-arrays all *JJs* are 3μm wide and are separated by superconducting rectangular loops of area A$_{loop}$=3×15 μm$^2$. A bias current *I* is applied symmetrically via the central top and bottom electrodes and *V* is measured across the array. A magnetic field *B* is applied perpendicular to the planar array's structure via a control current $I_{ctrl}$ through an inductively coupled coil. Consequently, a *dc* external magnetic flux, $\Phi_{dc}$, is coupled into the array. The *JJ*-array was coupled to the resonant modes of a millimeter wave Fabry-Perot resonator [30] with an estimated quality factor of 3500 [31] which was excited at a TEM00k microwaves resonance in the frequency range, $f_P=\omega_P/2\pi$, (45-75) GHz at an input power level up to 30 mW (see Fig. 1b). The coupling which was controlled by the rotation of the array relative to the electric field in the waveguide, is minimal when E is parallel to the grid and maximal in the perpendicular direction. The equivalent electrical circuit of the 100 JJ-array coupled to the Fabry-Perot resonator is shown in Fig. 1c. Families of *dc IV* curves (*IVCs*) for different values of *Φ* were measured for the arrays by a 4 point-contact method. The applied $\Phi_{dc}$ excites multiple flux-flow resonances in the array's IVCs, while the applied MW pump at f$_P$ induces integer Shapiro steps [32] at voltages V given by the *ac* Josephson equation $f_P$=nV/$\Phi_0$, a well as half-integer Shapiro steps [33, 34] at frequencies $f_P$=(n-1/2)V/$\Phi_0$, with n=1, 2, 3, ….

To observe robust, well-pronounced, B-field tunable flux-flow resonances the discreteness parameter $\beta = 2\pi L I_c/\Phi_0$ should be smaller than 2 [25-26]. Here $I_c$ is the *JJs* critical current and *L* is the inductance of the superconducting loops. On the other hand, to avoid flux-flow resonances interfering with other unwanted effects, such as the appearance of the hysteresis on the IVCs or/and of the *LC* geometric resonances (similarly to those observed in dc SQUIDs [35-37]), the Stewart-McCumber parameter $\beta_c = 2\pi I_c R^2 C/\Phi_0$ should be less than 1. Here $C$ and $R$ is the capacitance and the normal state resistance of the $JJs$, respectively. Since both $\beta$ and $\beta_c$ are proportional to $I_c$, the Josephson critical current density $J_c$ needs to be kept relatively low. 24° bicrystals used in our report have relatively large values for $J_c$. Considering this we had to employ medium-underdoped YBCO films characterized by relatively small values for both $J_c$ and its critical temperature $T_c$=49 K. For the $JJ$-array design/fabrication parameters used in our report the required conditions $\beta < 2$ and $\beta_c < 1$ could only be fulfilled in the range (14-45) K. To extend the range of operation temperatures of such devices from mK range to 90 K optimally doped YBCO films deposited on 30° symmetric bicrystals characterized by smaller values for $J_c$ should be used instead. Alternatively, other $JJ$ fabrication methods could be used such as ramp-type *JJs* characterized by a better control of $J_c$ values. Finally, thinner YBCO films could be used or *JJ* arrays with smaller values for *L*, *R* and *C* should be fabricated.

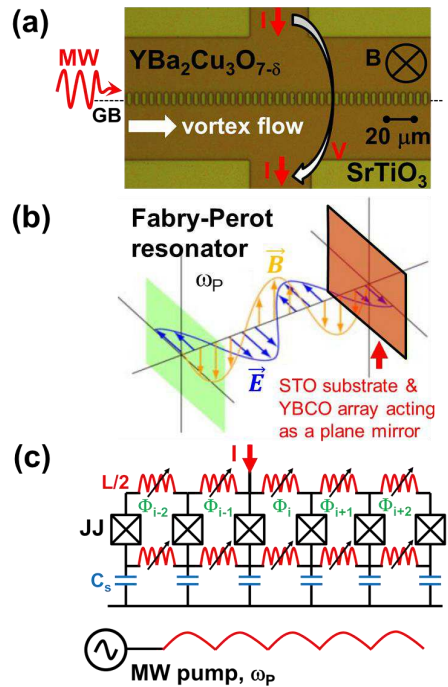

Fig. 1. a) Optical micrograph of the 100 JJ-array showing its central part (the middle 42 JJs). The JJs are formed across the bicrystal grain boundary (GB) shown with dotted line. A magnetic field $B$ is applied perpendicular to the planar structure. When biased with a dc current $I$ magnetic vortices start to flow across the 100 JJ-array in the direction show by the bold horizontal arrow. When vortices exit the array EM-radiation is emitted. b) Being inductively coupled to a Fabry-Perot resonator, the 100 JJ-array is parametrically pumped at a frequency $f_P = \omega_P/2\pi = 45$ GHz by modulating the flux threading the array's loops; c) Schematic view of the equivalent circuit. $C_s$ is the parasitic capacitance of the substrate.

In Fig. 2, 18 consecutive IVCs and their corresponding dI/dV(V) curves are plotted, respectively, showing both the flux-flow resonance m=1 and the first half-integer ½ and the first integer n=1 Shapiro steps. As expected, the voltage position $V_{res}$ of the flux-flow resonance does change with B, while for the Shapiro steps it does not. Flux-flow resonances as well as Shapiro steps are better defined on *dI/dV's* relative to the *IV's* data: they appear as peaks as opposed to current steps. For that reason, in the following the *dI/dV* data will be analysed. Typical full data sets recorded at two temperatures T=40 K and 45 K consisting of families of 161 IVCs (a 1,000 points each) measured for different values of $I_B$ changed in steps of 5 μA in the range (-0.40, 0.4) mA are shown in Figs. 3a, 3b, 4a and 4b. Figs 3a and 4a are for the case when no MW is applied, while Figs. 3b and 4b are for the case when the 100 *JJ*-array is parametrically pumped at a frequency $f_P = \omega_P/2\pi = 45$ GHz by modulating the flux threading the array's loops. The 3D plots of $dI/dV(V, I_B)$ shows a periodicity with $I_B$ of 0.235 mA that corresponds exactly to an additional $\Phi_0$ in each superconducting loop area $A_{loop}$ separating the $JJ$s as determined from calibration measurements of periodic voltage oscillations of YBCO-SQUIDs with identical loop areas patterned separately [29]. Such 3D maps of flux-flow resonances are very useful in representing the complexity/richness of flux-flow dynamics involving multiple flux-flow resonances m=1-4, their mutual interference/splitting and $\Phi_0$-periodicity with $B$. For positive current biases, resonance m=1 split into two, m=1 and m=2, at V=0.04 mV. Initially the voltage position of the m=1 resonance linearly increases with increasing $B$ up to V=1.6 mV, but then slightly decreases linearly with $B$ at the same rate (see also Fig. 2b). The voltage position of the m=2 resonance linearly decreases with increasing $B$. This cycle repeats itself with a $\Phi_0$-periodicity. As a result, a peculiar pattern of small and large triangles overlapping each other is formed. For negative current biases the pattern of the flux-flow resonance dynamics is similar but lacks both reflection symmetry with respect to V=0 axis and point symmetry with respect to $B$=0 as it is shifted on the $B$ field axis by a value corresponding to a fraction of $\Phi_0$. This shift is due to the presence of a remnant non-zero flux associated with the unshielded earth magnetic field in the array/cryostat. The height of m=1, 2 resonances, and consequently, the power of emitted EM-radiation, both monotonically decrease with increasing voltage. As $V_{res}$ of various flux-flow resonances changes with $B$ in the range (-0.05, 0.05) mV the frequency, $f$=V/$\Phi_0$, of the emitted EM-radiation changes in the range (0, 24) GHz.

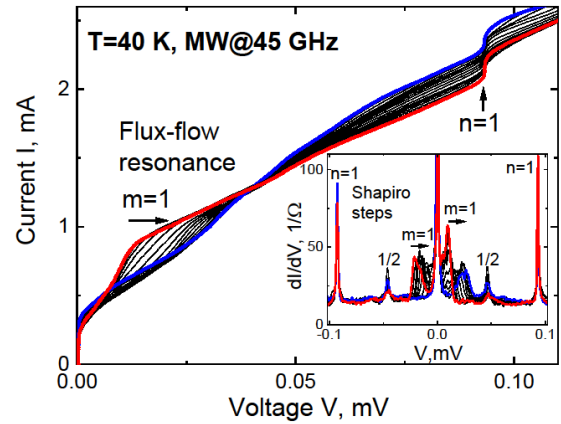

Fig. 2. Experimental data set at a temperature of 40K and applied MW pump at $f_P$=45 GHz. a) 18 consecutive IVC's for positive voltages of the 100 JJ-array measured for different applied B fields (or equivalently, fluxes $\Phi$) with $I_B$ changed in steps of 5 μA showing the flux-flow resonance m=1 and the first half integer ½ and integer n=1 Shapiro steps. Inset: dI/dV(V) for both negative voltages and positive voltages. The horizontal arrows show the shifts of the peak voltage $V_{res}$ of m=1 with B. The central peak around V=0 is due to the array's Josephson critical current.

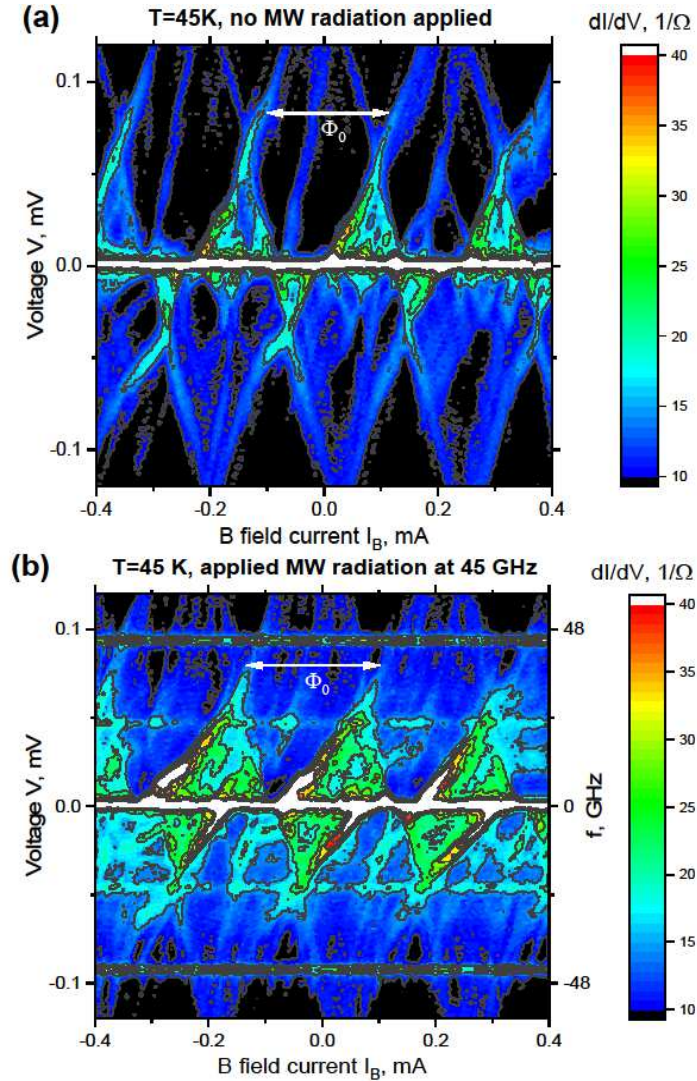

Fig. 3. 3D plots of dI/dV(V, $I_B$) at T=45 K extracted from a family of 160 IVC's measured for a 100 JJ-array for different values of the B field current $I_B$. $I_B$ was changed in steps of 5 μA in the range (-0.4, 0.4) mA. (a) is for the case when no MW is applied and (b) is with an applied MW pump at $f_P$=45 GHz. The rich network of non-horizontal branches crossing each other and showing a $\Phi_0$-periodicity are due to multiple flux-flow resonances. The white horizontal branches centred at V=0 are due to the array's Josephson junction's critical current, while the horizontal branches just below ±0.1 mV in (b) are due to the first Shapiro steps n=1.

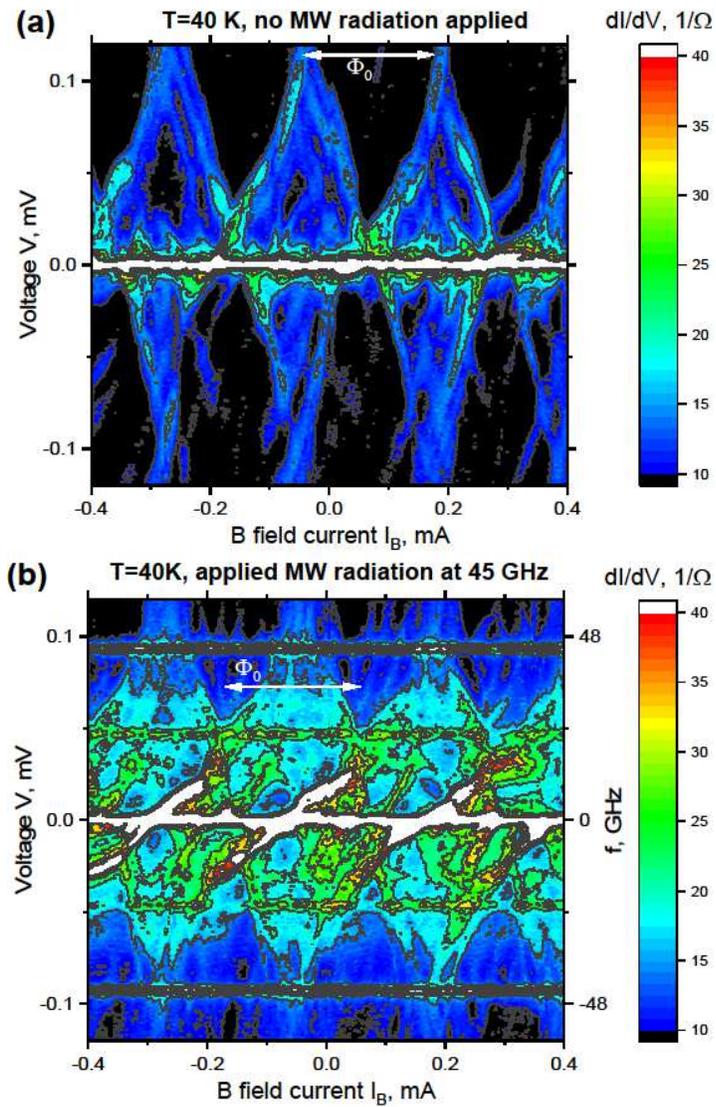

Fig. 4. 3D plots of dI/dV(V, $I_B$) as described in Fig. 3 but recorded at 40 K.

The most remarkable feature observed experimentally is the increase in the height of the flux-flow peaks recorded in Figs. 3b and 4b (recorded with applied MW) relative to Figs. 3a, and 4a, (recorded with no applied MW), respectively. Thus, in particular, with applied *MW* the flux-flow branches include significant white colour regions (corresponding to values above 40, see Figs, 3b and 4b). An increase in the height of the flux-flow peak may be produced by simple synchronization of the oscillator currents by the pump as in Fabry-Perot resonant optical feedback of lasers [38, 39]. In this case, considering that the linewidth of the MW pump is narrow (as evidenced by the small width of Shapiro steps), it follows that the linewidth of the flux-flow-oscillator should decrease, leading to a seemingly higher peak. This is in high contrast to what we observe in our experiments. Indeed, the application of MW results in an increase of the height of flux-flow resonances only, while their width remains unchanged. This is clearly visible in the 3D plots of Figs. 3 and 4 where the width of flux-flow branches does not change with the application of MW, as well as in the detailed 2D plot of Fig. 5. The increase

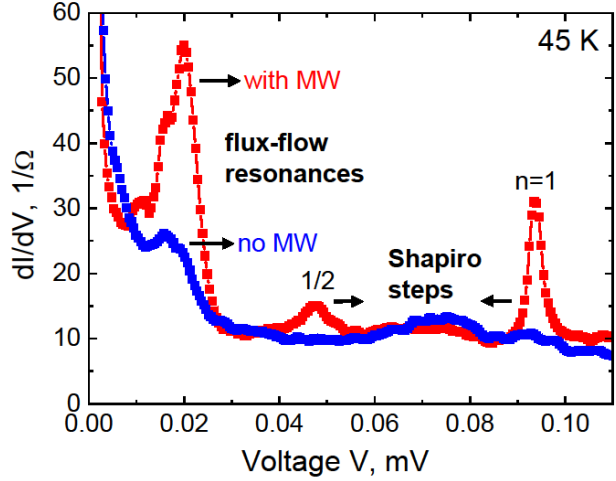

Fig.5 dI/dV(V) showing the impact of MW pump at 45 GHz on the flux-flow resonances: their height is significantly increased with their width unchanged, consistent with an amplification due to the JJ-array's inductance being parametrically pumped.

in the height of flux-flow resonances does not occur at the expense of their width which strongly suggests that the observed effect is due to parametric amplification.

Parametric modulation (pumping) of the superconducting loop "i" (i takes values between 1-100) inductances $L_i(t)$ (at higher rate than the natural frequency) can inject energy into the circuit if the pump phase is set to increase the inductance when the current through the SQUID inductor is at its maximum and minimum. This can be understood as a modulation of the loop magnetic potential energy, $U_B(t) = I_{circ,i}^2 L_i(t)/2$ where increasing the inductance at the current extrema corresponds to work, generating even more current. Here $I_{circ,i}$ is the circulating current in loop "i" and $L_i(t)$ has two parametrically modulated components, the Josephson junction inductance $L_{JJ}(t)$ [40], and the SQUID inductance $L_{SQUID}(t)$ [11]:

$$L_i(t) = 2L_{JJ}(t) + L_{SQUID}(t) = \frac{\Phi_o}{I_c}\left[1 + \frac{I^2(t)}{2I_c^2}\right] + \Phi_0\left|2I_c\cos\left(\pi\frac{\Phi_{app}}{\Phi_0}\right)\right|^{-1}, (1)$$

Here $I(t) = I_0\cos(\omega_p t)$ is the small alternating current with amplitude $I_0$ induced by the pump and $\Phi_{app}$ is the applied flux to each loop "i" in the array $\Phi_{app}(t) = \Phi_{dc} + \Phi_{ac}(t) = \Phi_{dc} + \Phi_{ac,0}\cos(\omega_p t)$ with $\Phi_{ac,0}$ being the amplitude of the ac applied flux component. The MW pump induces two significant effects on the flux-flow resonances. Firstly, the intensity and dynamics of flux-flow resonances is dramatically enhanced when MW is applied via parametric amplification due to pumping of the inductances $L_i(t)$. As we move along the main flux-flow resonance branch in the ($V$, $B$) plane the parametrically-induced increase of the height of the flux-flow peak was observed for all temperatures measured. Secondly, the frequency of the flux-flow induced EM increases faster with B when MW is applied relative to the case when no MW is applied. Indeed, the slope of the flux-flow branches relative to the horizontal axis (V=0) in the 2D plane (V, B) is significantly reduced when MW is applied (compare Figs. 3a with 3b and 4a with 4b). Interestingly, the temperature dependence of the amplitude of flux-flow resonances with or without MW differ from one another. Thus, when no MW is applied when decreasing T from 45 K to 14 K the height of flux-flow resonances decreases monotonically with the height of the flux-flow peak decreasing by a factor of 2 on average between 45K and 40 K (see Figs 3a and 4a) and, subsequently, a factor of 2.5 on average between 40K and 30 K. In contrast, however, when MW is applied, the intensity of flux-flow resonances first increases when temperature is reduced from 45K to 40 K (compare Fig. 3b and 4b) and then decreases when T is lowered further to 30K. At 30K the height of the flux-flow peak is on average 2 times lower than at 45K. This strongly suggests that the gain of the parametric amplification increases with temperature. To explain all these features, as well as to fully understand the mechanism of parametric amplification in our devices, requires an extension of the previous theoretical investigations of similar arrays [23-27, 29, 41, 42] to include the excitation of *MW* modes in such 1D Josephson transmission lines coupled to a Febry-Perot resonator. Since frequency and voltage are related by the Josephson relation $f=V/\Phi_0$ we were able to calculate the parametric amplification gain versus frequency from the ratio of flux-flow resonance peak height (with and without MW), versus voltage. Thus, the associated parametric gain was in the range (6, 7.8) dB at 45 K, (8, 10.4) dB at 40 K, and (6.5, 8.1) dB at 30 K. Remarkably, at all temperatures measured the gain does not change significantly in the entire frequency range (1-23) GHz (see for example Fig. 6) were flux-flow resonances were observed suggesting a wide bandwidth feature for the parametric amplification. For temperatures in the range (14-30) K the magnetic field tunability of flux-flow resonances is strongly suppressed with a parametric gain between (5-6) dB observed for a reduced range of signal frequencies (1-5) GHz. Below 14 K the flux-flow resonances could not be resolved due to $\beta$ and/or $\beta_c$ increasing above the threshold values 2 and 1, respectively, required for their unambiguous interpretation [24, 25, 35-37]. Consequently, the strength of parametric amplification could no longer be tested below 14 K. This does not imply that the device does not parametrically amplify below 14 K. It is important to make an estimation of the *EM* power $P_{PA}$ received/absorbed from the pump at 45 GHz that is responsible for the observed parametric amplification (*PA*) of flux-flow resonances. This can be done by realizing that the same absorbed *MW* power is responsible for the excitation of Shapiro steps. Consequently, the following analytical formula can be implemented [43]:

$$P_{PA} = P_{Shapiro} = \frac{1}{2}\left(4I_{supp}I_c\frac{\omega_p^2}{\omega_c^2}\right)R_{N,array}, \quad (2)$$

where $I_{supp}$ is the suppression of the array's critical current (the difference in the critical currents in the *IVCs* with and without MW), $\omega_r$ is the array's characteristic frequency, and $R_{N, array}$ is the array's normal resistance. A value of 10 nW is obtained using Eq.(2).

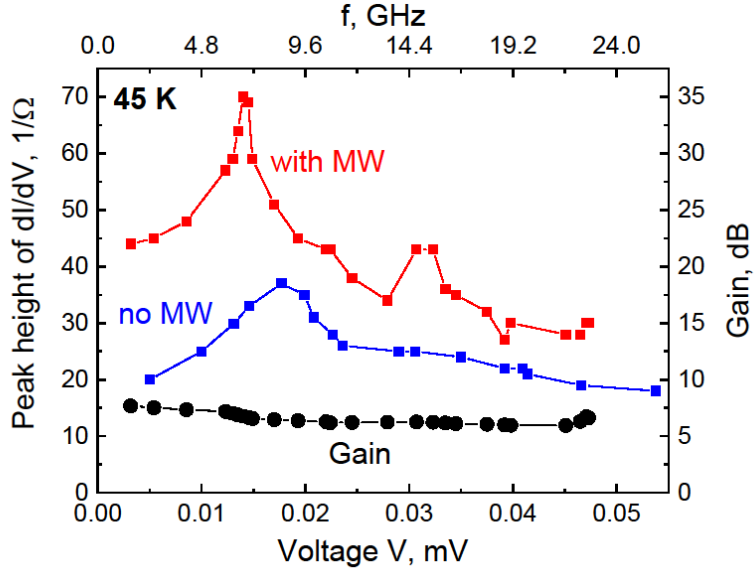

*Fig.6. Voltage (frequency) dependence of the peak height of dI/dV of the flux-flow resonance at 45 K, with or without applied MW at 45 GHz. Parametric gain (right axis) versus frequency.*

We observe parametric amplification of *EM* produced by flux-flow oscillators made of YBCO Josephson junctions arrays coupled to the resonant modes of a millimeter wave Fabry-Perot resonator at a pump frequency of $\omega_p$ =45 GHz. A significant parametric gain was measured for signal frequencies $\omega_S$ in the range (1, 25 GHz) and temperatures between 30 K and 45 K with a maximum of (8-10.4) dB reached at 40K. In the temperature range (14-30) K a parametric gain between (5-6) dB was observed for a reduced range of signal frequencies (1-

5) GHz. With an appropriate adjustment of design/fabrication parameters such tunable *MW* generators/detectors and parametric amplifiers made of YBCO could potentially operate in a much wider range of temperatures (10 mK-77K). The power $P_{PA}$ absorbed from the pump was estimated to be about 10 nW. Increasing $P_{PA}$ by implementing more efficient or alternative coupling mechanisms [14-22] should result in an increase of the parametric gain. Although in our case the *MW* signal of frequency $\omega_S$ to be amplified is intrinsically generated by a chain of Josephson vortices moving unidirectionally, in principle, the concept of parametric amplification by the *JJ*-array demonstrated here should work for externally applied EM signals as well. To fully understand the mechanism of parametric amplification in such 1-D Josephson transmission lines coupled to a Fabry-Perot resonator and their potential for applications a theoretically investigation at this stage would be highly desirable.

Data availability. The data that support the findings of this study are available from the corresponding author upon reasonable request.

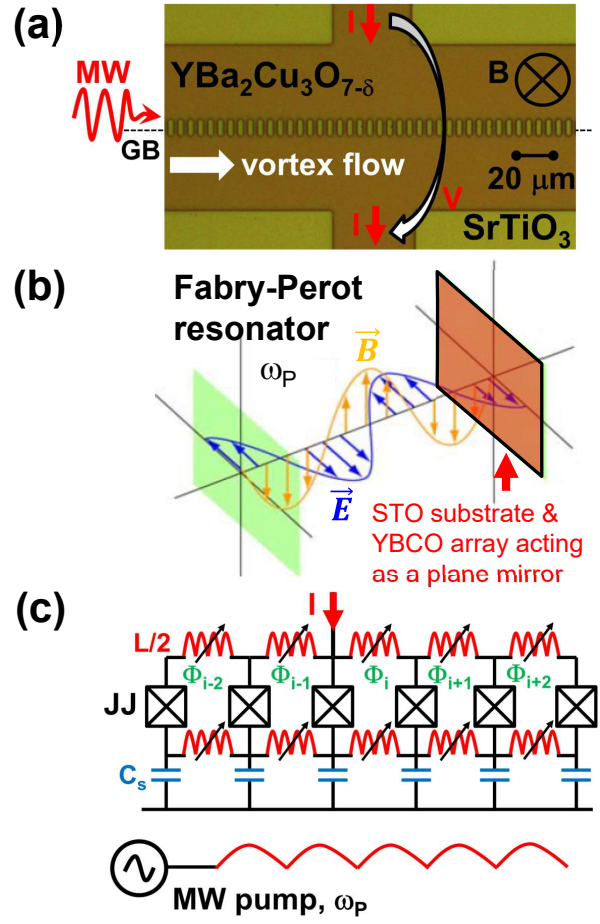

**(a)** MW — GB — YBa$_2$Cu$_3$O$_{7-\delta}$ — I — B ⊗ — vortex flow — 20 μm — V — SrTiO$_3$

**(b)** **Fabry-Perot resonator** — $\omega_P$ — $\vec{B}$ — $\vec{E}$ — STO substrate & YBCO array acting as a plane mirror

**(c)** L/2 — I — JJ — $\Phi_{i-2}$ — $\Phi_{i-1}$ — $\Phi_i$ — $\Phi_{i+1}$ — $\Phi_{i+2}$ — C$_s$ — MW pump, $\omega_P$

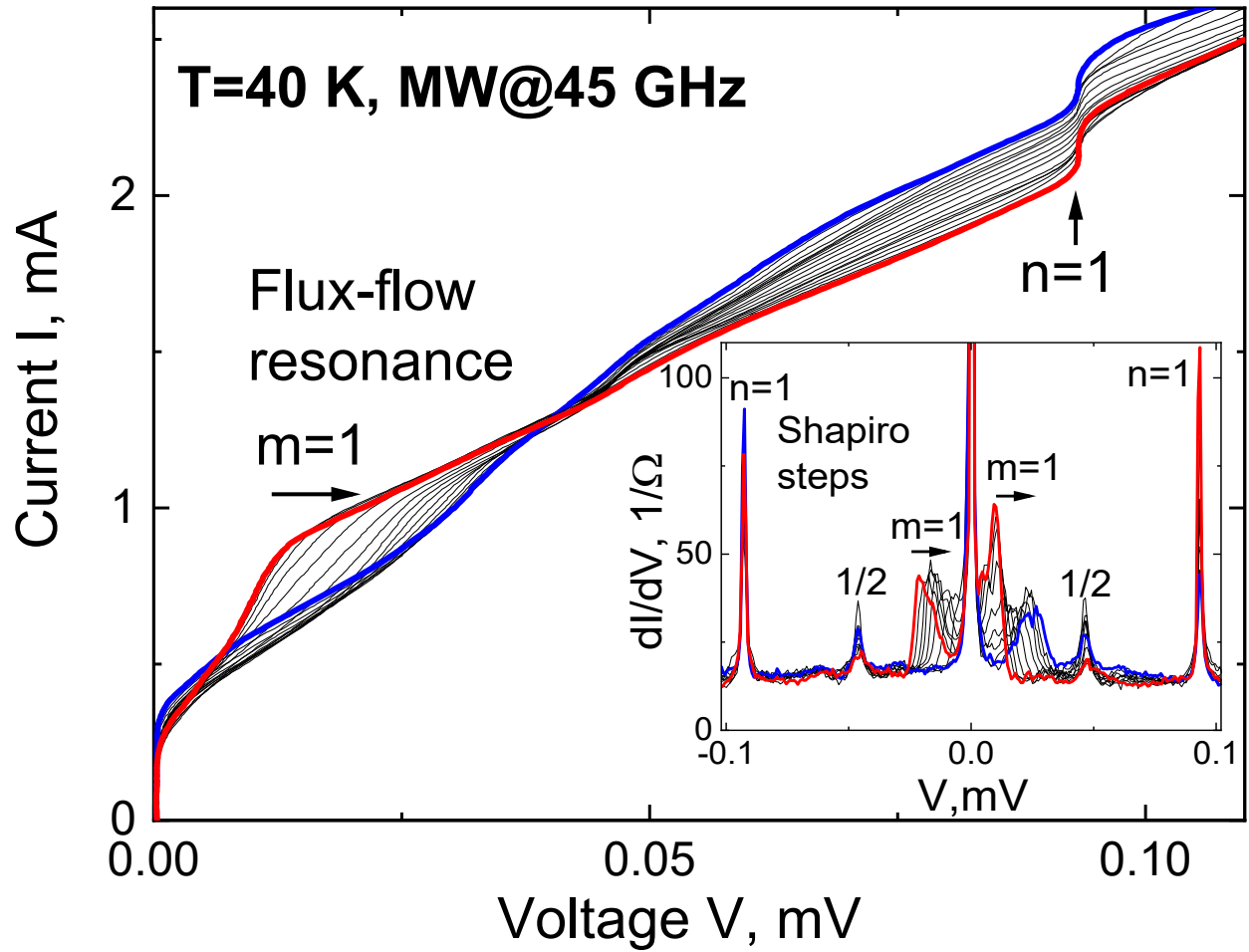

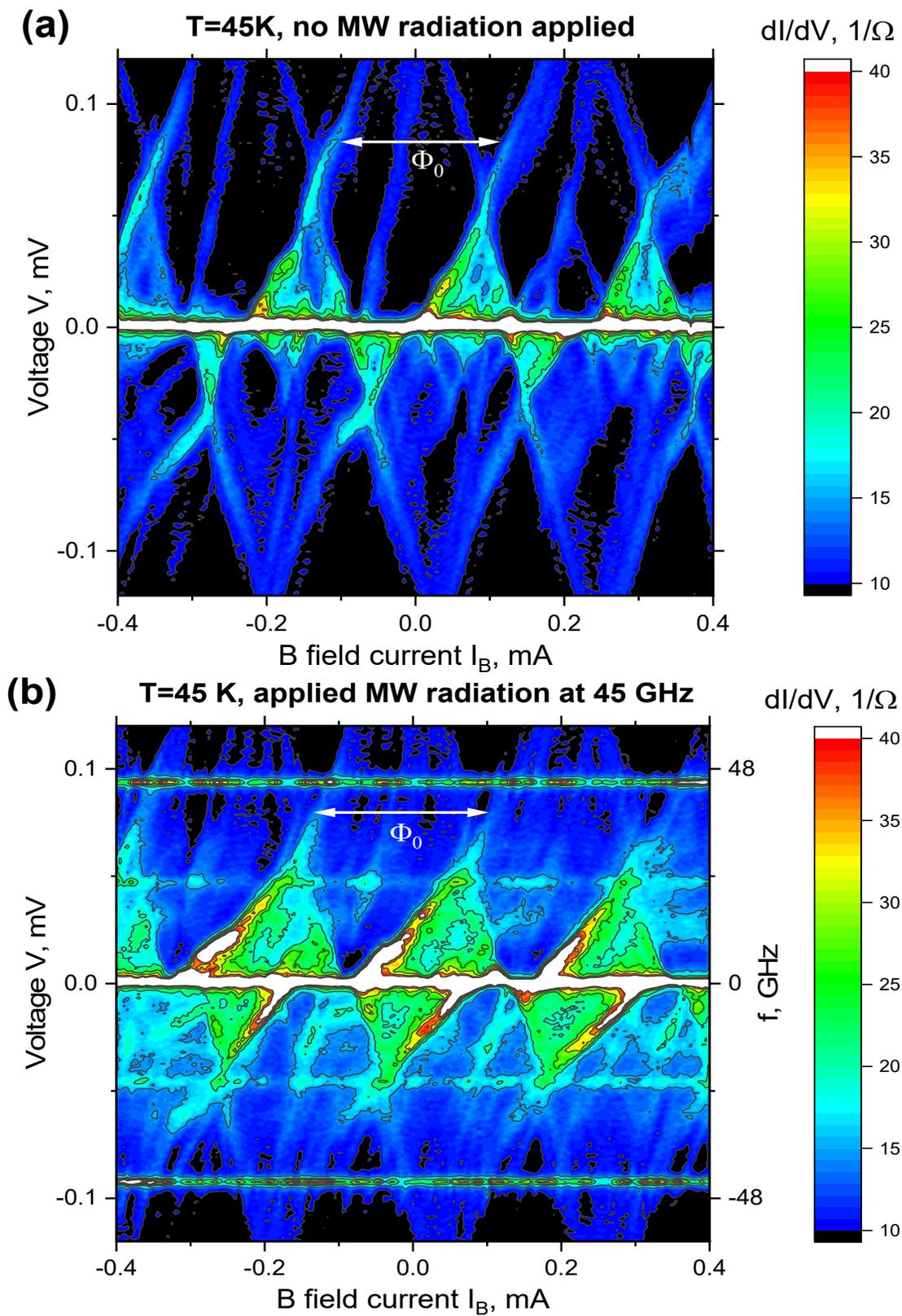

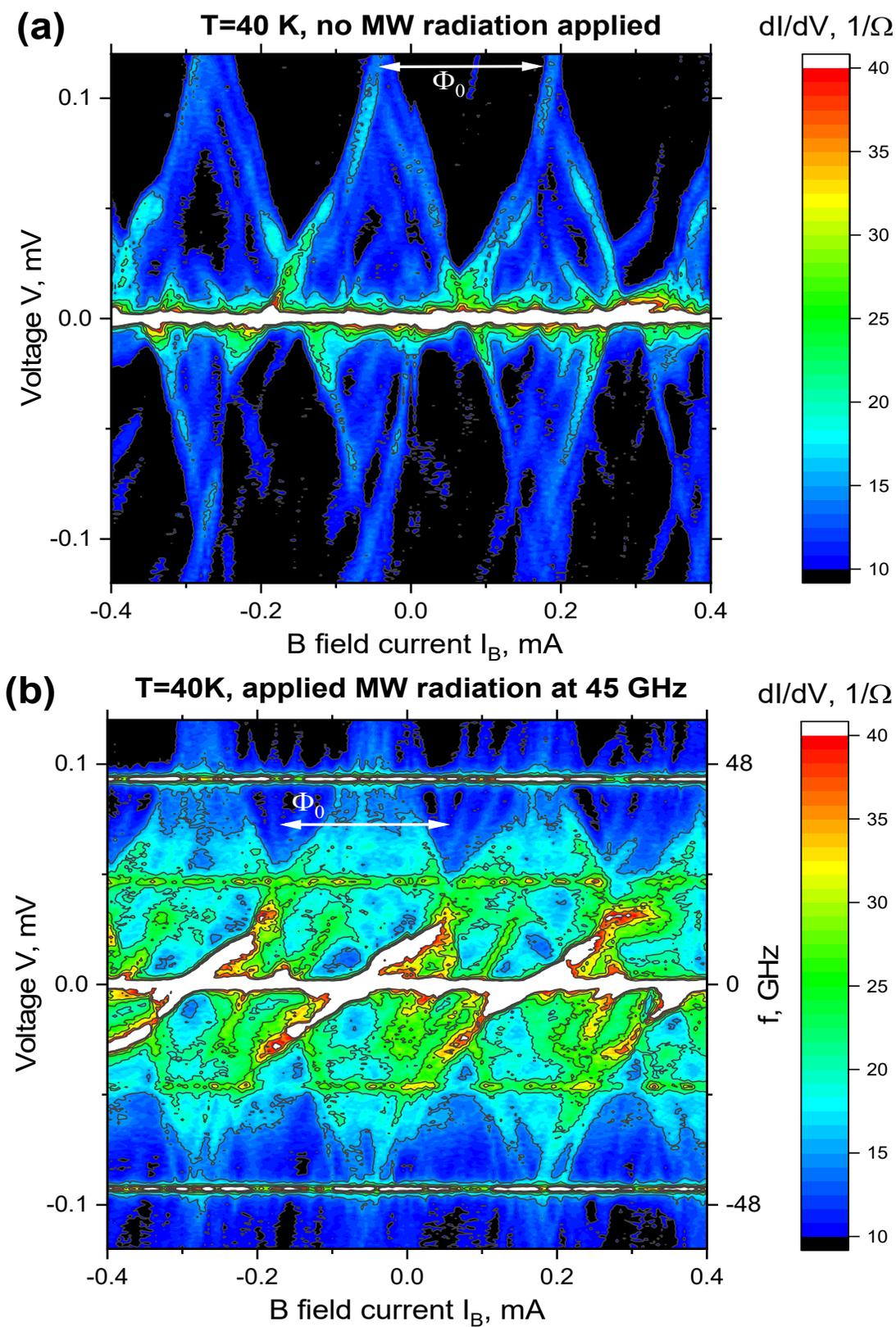

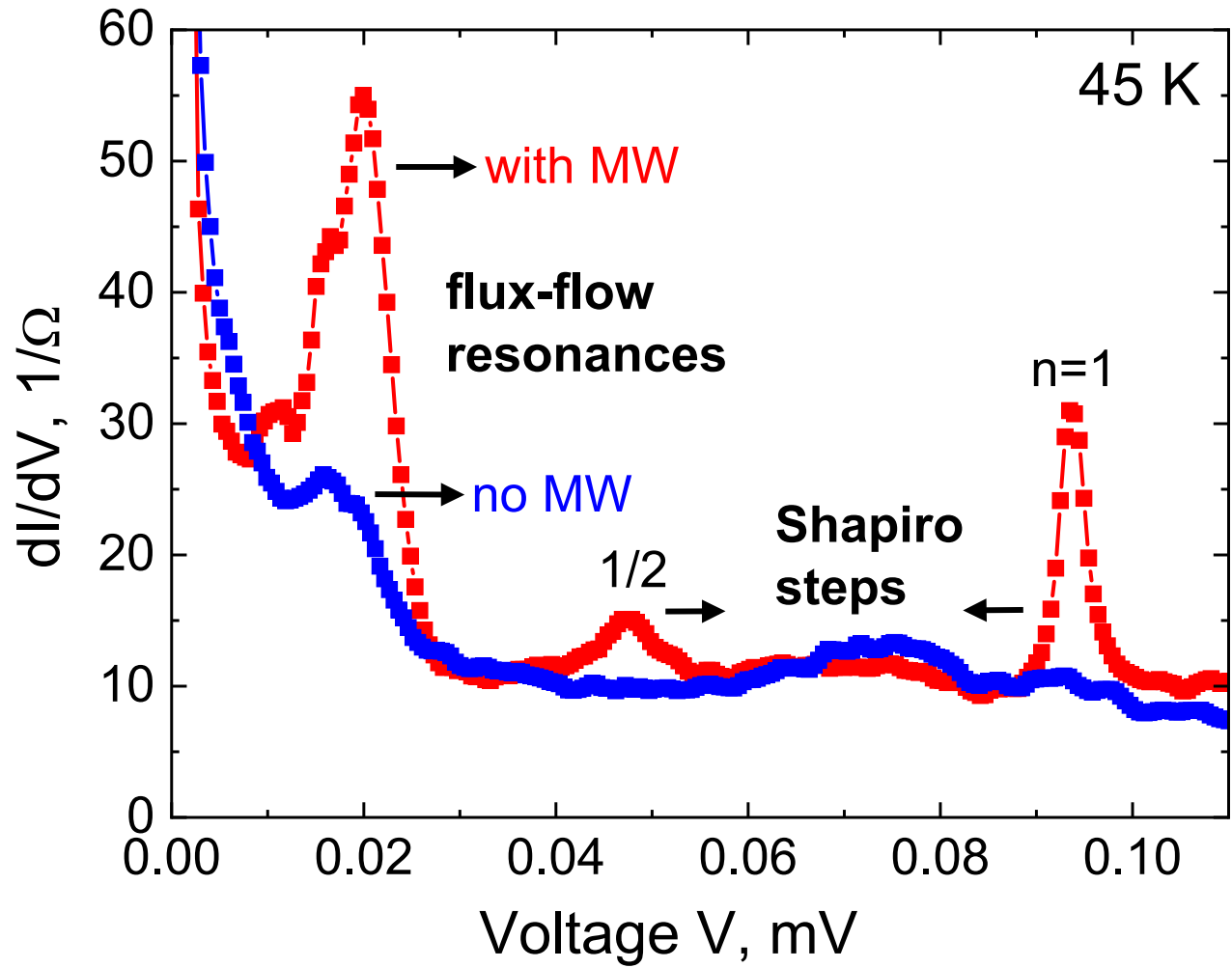

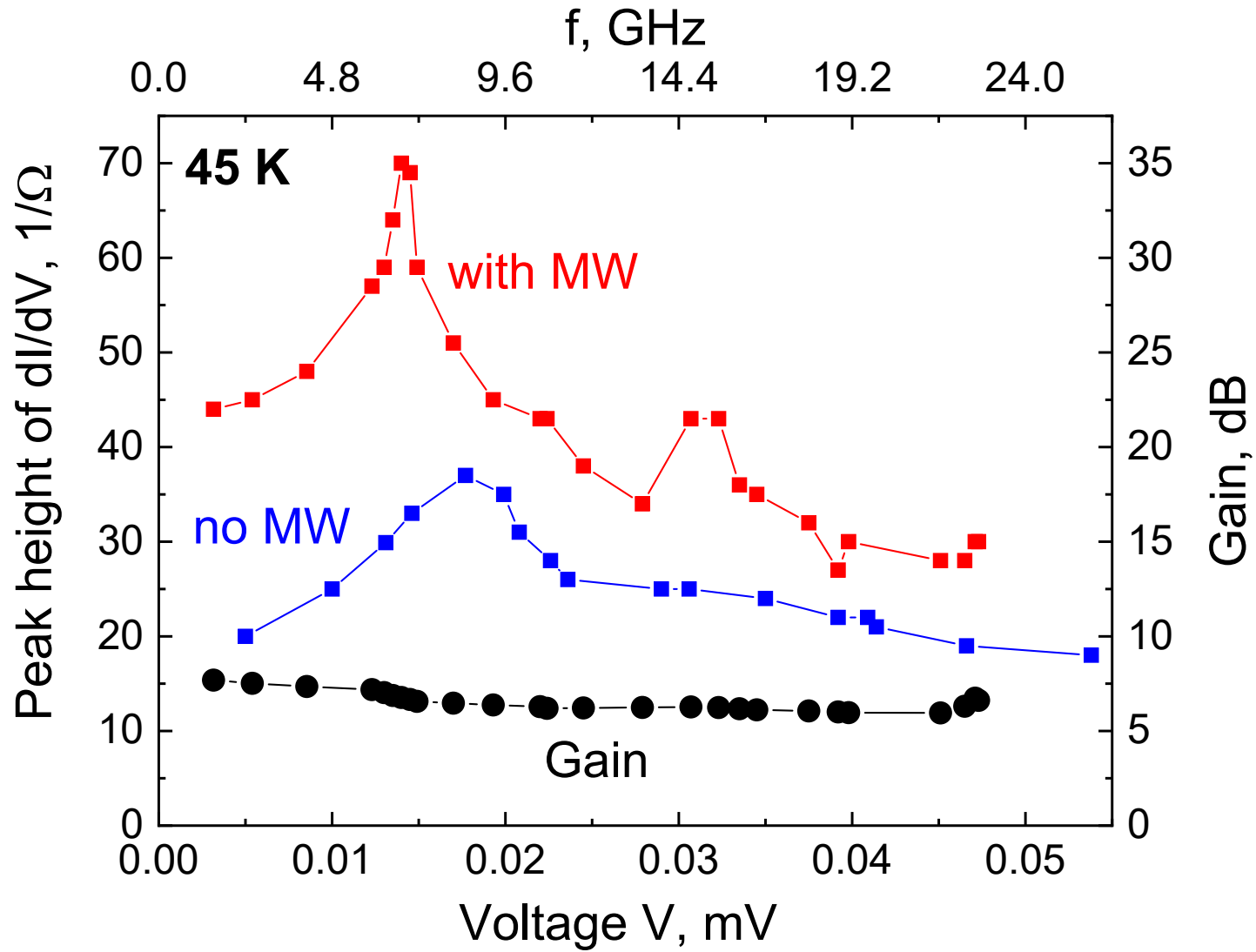